%% file: _Head__Dissipation_short-paper.tex
\documentclass[english,prl,twocolumn,superscriptaddress]{revtex4-2}

\usepackage[T1]{fontenc}
\usepackage[utf8]{inputenc}
\usepackage{xcolor}
\usepackage{babel}
\usepackage{verbatim}
\usepackage{refstyle}
\usepackage{amsmath}
\usepackage{amssymb}
\usepackage{graphicx}
\usepackage{wasysym}
\usepackage{esint}
\usepackage{microtype}
\usepackage[pdfusetitle,
 bookmarks=true,bookmarksnumbered=false,bookmarksopen=false,
 breaklinks=false,pdfborder={0 0 0},pdfborderstyle={},backref=false,colorlinks=true]
 {hyperref}
\hypersetup{
 urlcolor=blue}

\makeatletter

\AtBeginDocument{\providecommand\appref[1]{\ref{app:#1}}}

\usepackage{chngcntr} 

\newref{fig}{ refcmd={Fig.~\ref{#1}} }
\newref{tab}{ refcmd={Table~\ref{#1}} }
\newref{eq}{ refcmd={(\ref{#1})} }
\newref{subsec}{ refcmd={subsection~\ref{#1}} }
\newref{subsecapp}{ refcmd={subsection~\ref{#1}} }
\newref{sec}{ refcmd={section~\ref{#1}} }
\newref{app}{ refcmd={the End Matter} }
\newref{part}{ refcmd={Part~\ref{#1}} }

\makeatother

\begin{document}
\input{_Text__Title.tex}

\input{_Text__Intro.tex}

\input{_Text__Dissipative-response.tex}

\input{_Text__Discussion.tex}

\bibliographystyle{apsrev4-2_improved.bst}
\bibliography{Bibliography.bib}

\appendix
\input{_App__End-Matter.tex}

\end{document}

%% file: _Text__Title.tex
\title{Dissipation due to Bulk Localized Low-Energy Modes in Strongly Disordered
Superconductors}
\author{Anton V. Khvalyuk}
\email{anton.khvalyuk@protonmail.com}

\affiliation{LPMMC, Université Grenoble Alpes, 38000 Grenoble, France}
\author{Mikhail V. Feigel'man}
\email{mikhail.feigelman@nanocenter.si}

\affiliation{Nanocenter CENN and Jozef Stefan Institute, Jamova cesta 39, Ljubljana
1000, Slovenia}
\date{\today{}}
\begin{abstract}
Strongly disordered superconductors (SDSCs) are widely used in qubits,
microwave resonators, photon detectors, and other superconducting
quantum devices. In SDSC-based devices, coherence times are limited
by low-temperature microwave dissipation in the material. However,
the standard Mattis-Bardeen theory fails in SDSCs because their single-particle
spectrum exhibits a hard pseudogap~$\Delta_{P}$ both below and above
the transition temperature~$T_{c}$. We develop a novel microscopic
theory of the dependence of ac~dissipation in such systems on temperature~$T$
and frequency~$\omega$. We analyze the resonator quality factor
$Q(\omega,T)$ in the practically relevant range $\hbar\omega,\,T\ll\Delta\leq\Delta_{P}$,
where $\Delta$ is the typical superconducting order parameter, distinct
from~$\Delta_{P}$. We show that low-$\omega$ dissipation is dominated
by a new type of bulk localized collective modes arising from spatial
inhomogeneity of the superconducting state. Consequently, $Q(\omega)$
decreases strongly with~$\omega$ and exhibits two-level-system-like
growth with~$T$ for $T\ll T_{c}$. Our theory provides a microscopic
understanding of existing and future experiments on thin films of
$\mathrm{InO}_{x}$, TiN, NbN, and similar SDSCs, and is phenomenologically
relevant to granular aluminum films. The results suggest strategies
to mitigate intrinsic microwave losses in SDSC-based quantum devices.
\end{abstract}
\maketitle

%% file: _Text__Intro.tex
\paragraph{Introduction.\label{sec:Introduction}}

Superconducting quantum circuits rely on a simple promise: at microwave
frequencies and at temperatures well below the superconducting gap,
a superconductor should behave as an almost lossless inductor. In
conventional $s$-wave superconductors this expectation is formalized
by Mattis-Bardeen theory~\citep{MattisBardeen}, which predicts an
exponentially small dissipative conductivity, $\text{Re}\sigma$,
while disorder increases the sheet kinetic inductance~$L_{K}$. This
makes disordered films a powerful route to high-impedance quantum
circuits, compact resonators, enhanced zero-point voltage fluctuations,
and protected-qubit or detector architectures~\citep{Mooij05,kermanMetastableSuperconductingQubit2010,Doucot12,Kitaev13,peltonenHybridRfSQUID2018,Groszowski18,Mooij16,Grunhaupt2018,Bylander19,Mooij06,Astafiev12,DeGraaf18,Astafiev22}.
The same route, however, exposes a central materials problem: increasing
$L_{K}$ is often accompanied by an excess of conductive loss, as
seen in a recent survey of the microwave quality factor~$Q$ of superconducting
quantum devices~\citep{charpentierUniversalScalingMicrowave2025}.
Understanding the microscopic origin of this intrinsic loss is therefore
essential both for the physics of disordered superconductivity and
for the engineering of coherent quantum circuits.

Strongly disordered superconductors (SDSCs) are not merely dirty versions
of ordinary BCS metals~\citep{Feigelman_Fractal-SC_2010,SFK-review-2020}.
Close to the superconductor-insulator transition (SIT), amorphous
films such as $\mathrm{InO}_{x}$, TiN, and NbN develop a hard single-particle
pseudogap $\Delta_{P}$ that can persist well above the superconducting
transition temperature~$T_{c}$~\citep{SFK-review-2020}. This
pseudogap signals the formation of spatially localized Cooper pairs
before global phase coherence is established. Below $T_{c}$, the
superconducting order parameter is itself strongly inhomogeneous~\citep{Sacepe2008,Sacepe_2011_for-pair-preformation,Dubouchet_2018_Preformation-of-Pairs}:
its local magnitude has a broad distribution, with rare regions where
the gap is much smaller than its typical value~\citep{Feigelman_Fractal-SC_2010,op-distribution-paper,Feigelman_SIT_2010}.
As a result, the low-energy electromagnetic response of an SDSC cannot
be inferred by simply adding disorder to the Mattis-Bardeen quasiparticle
picture. The relevant low-energy degrees of freedom may instead live
inside the Cooper-pair sector of an inhomogeneous condensate.

This distinction is sharpened by recent microwave experiments~\citep{charpentierFirstorderQuantumBreakdown2025}.
Thin films of amorphous $\mathrm{InO}_{x}$ can reach extremely large
kinetic inductance, up to $L_{K}\approx17\,\mathrm{nH}/\square$,
but at the cost of a strongly suppressed resonator quality factor~\citep{Charpentier2023_thesis,charpentierFirstorderQuantumBreakdown2025,charpentierUniversalScalingMicrowave2025}.
The usual extrinsic explanations are unsatisfactory in this regime:
surface dielectric loss is inconsistent with the weak dependence on
electric-field participation ratio~\citep{Charpentier2023_thesis},
atomic two-level systems have no natural reason to track the electronic
disorder so sharply, and conventional thermal or nonequilibrium quasiparticles
cannot account for the magnitude and disorder dependence of the loss
in the presence of a large pseudogap~\citep{Sacepe_2011_for-pair-preformation,Feigelman_Fractal-SC_2010,charpentierUniversalScalingMicrowave2025}.
The basic unresolved question is therefore simple and experimentally
pressing: what microscopic objects absorb microwave photons with $\hbar\omega,\,T\ll\Delta\leq\Delta_{P}$
in a pseudogapped superconductor?

Previous theoretical work on SDSCs established the static ingredients
needed to address this question: a pseudospin description of localized
preformed Cooper pairs, a broad distribution of the local order parameter,
and rare ``weak spots'' that strongly affect the temperature dependence
of the superfluid stiffness~\citep{Feigelman_Fractal-SC_2010,op-distribution-paper,superfluid-density-paper-2024}.
What remained missing was the dynamical step: identifying the finite-frequency
modes of this inhomogeneous condensate, computing their contribution
to $\text{Re}\sigma(\omega,T)$, and relating the result directly
to the measured resonator quality factor~$Q$. In this Letter, we
perform this step. We show that rare weak spots host localized collective
modes corresponding to low-energy rearrangements of Cooper pairs,
with an electric dipole moment set by the weak-spot size rather than
by pair breaking. These modes produce a bulk dissipative response
with a two-level-system-like factor $\tanh(\hbar\omega/2T)$ and a
strong frequency dependence controlled by the low-value tail of the
order-parameter distribution~$P(\Delta)$, leading to a rapid drop
of~$Q$ with increasing~$\omega$. The resulting expression for
$Q(\omega,T)$ explains the main trends observed in $\mathrm{InO}_{x}$
resonators and turns microwave spectroscopy into a probe of the order-parameter
distribution of the rare weak regions that control dissipation in
SDSCs.

%% file: _Text__Dissipative-response.tex
\paragraph{A microscopic model of a SDSC.\label{sec:microscopic-description}}

The starting point of the microscopic model is the pseudospin Hamiltonian
describing localized preformed Cooper pairs that experience phonon-induced
attraction in the Cooper channel~\citep{Ma_Lee_1985_Ref-to-pseudospins,Ghosal2001,Feigelman_SIT_2010,Feigelman_Fractal-SC_2010}:
\begin{align}
H & =-\sum_{j}(\xi_{j}+e\phi_{j})2S^{z}_{j}\nonumber \\
 & -\sum_{jk}D_{jk}\left[S^{+}_{j}S^{-}_{k}e^{-i(2e/c)A_{j\rightarrow k}}+S^{-}_{j}S^{+}_{k}e^{i(2e/c)A_{j\rightarrow k}}\right].
\label{eq:pseudospin-hamiltonian}
\end{align}
Here, $j,k$ enumerate Anderson-localized single-particle states,
with state~$j$ characterized by energy $\xi_{j}$ and wave function
$\psi_{j}(\boldsymbol{r})$; $D_{jk}=\intop d^{3}\boldsymbol{r}\,D\left(\xi_{k}-\xi_{j}\right)|\psi_{k}(\boldsymbol{r})|^{2}|\psi_{j}(\boldsymbol{r})|^{2}$
is the matrix element of the local Cooper attraction, $D(\omega;\boldsymbol{r},\boldsymbol{r}')\approx D(\omega)\,\delta(\boldsymbol{r}-\boldsymbol{r}')$.
The pseudospin operators $S^{z}_{i},\,S^{\pm}_{i}$ provide a compact
encoding~\footnote{In terms of original electronic operators: $S^{-}_{j}=c_{j,\downarrow}c_{j,\uparrow}$,
$S^{+}_{j}=c^{+}_{j,\uparrow}c^{+}_{j,\downarrow}$, $S^{z}_{j}=c^{+}_{j,\uparrow}c_{j,\uparrow}+c^{+}_{j,\downarrow}c_{j,\downarrow}-\frac{1}{2}$.} of the absence of unpaired electrons at low temperatures due to a
large pseudogap~\citep{Feigelman_Fractal-SC_2010,Feigelman_SIT_2010}.
The interaction term in Eq.~(\ref{eq:pseudospin-hamiltonian}) therefore
induces tunneling of preformed Cooper pairs between different localized
states.

In what follows, we assume $\xi_{i}$ to be independent random variables
distributed according to a broad distribution~$P_{\xi}(\xi)$, with
a finite density at the Fermi level, $P_{0}:=P_{\xi}(\xi=0)=\nu_{0}/n$,
where $\nu_{0}$ is the single-particle density of states (DoS) per
spin projection, and $n$ is the electron concentration. The set of
sites $i$ and of pairs $\langle ij\rangle$ for which $D_{ij}>0$
can then be viewed as an \emph{interaction graph}. Because of strong
statistical fluctuations of $D_{ij}$, this graph is sparse~\citep{Feigelman_Fractal-SC_2010,superfluid-density-paper-2024}.
As a result, the immediate vicinity of each vertex is locally tree-like
with a certain average branching number $K$, whereas at large scales,
loops inevitably appear as a consequence of the embedding of this
graph into 3D Euclidean space. However, these loops almost surely
contain at least $m_{\text{tree}}\sim\ln\left\{ 2\nu_{0}r^{3}_{\text{loc}}\omega_{D}\right\} /\ln K\gg1$
sites, where $r_{\text{loc}}$ is the localization length of the
electron wave functions $\psi_{i}(\boldsymbol{r})$, and $\omega_{D}$
is the energy cutoff of the Cooper pair attraction. The quantity $m_{\text{tree}}$
thus describes the spatial extent of the locally tree-like structure.

Although the statistical distribution of $D_{ij}$ in a real system
is likely rather nontrivial~\citep{Feigelman_Fractal-SC_2010}, we
restrict ourselves to the following simple model: for a given $i$,
$D_{ij}=0$ for all $j$ except $K+1$ randomly selected neighbors
within the localization volume with equal probability, for which
$D_{ij}=\text{const}=\lambda/(2P_{0}K)$. This relation also defines
the dimensionless Cooper-pair coupling constant~$\lambda$. This
approximation is expected~\citep{op-distribution-paper} to be qualitatively
correct for low-energy physics unless the true $D_{ij}$~distribution
is broad.

The mean-field treatment of Hamiltonian~(\ref{eq:pseudospin-hamiltonian})
defines~\citep{op-distribution-paper} the superconducting energy
scale $\Delta_{0}\sim2\omega_{D}e^{-1/\lambda}$, where, in a more
realistic model of $D_{ij}$~\citep{Feigelman_Fractal-SC_2010},
the exponential changes to a power\nobreakdash-law dependence, $\Delta_{0}\propto\lambda^{1/\gamma}$
with~$\gamma\approx0.57$. Although the true order parameter is strongly
inhomogeneous at the scale of the coherence length~\citep{Sacepe_2011_for-pair-preformation,Ghosal2001,op-distribution-paper},
$\Delta_{0}$ provides a relevant energy scale. In particular, it
allows one to define the dimensionless disorder strength $\kappa=\overline{D_{ij}}/\Delta_{0}$,
which turns out to be the key measure of the competition between disorder
and superconductivity~\citep{op-distribution-paper,superfluid-density-paper-2024}:
$\kappa\ll1$ corresponds to a nearly homogeneous superconducting
state, whereas for $\kappa\gg1$ the model exhibits a broad distribution
of the order parameter that becomes fat tailed when $\kappa\ge\kappa_{1}=\exp\left\{ 1/2\lambda\right\} \gg1$,
with the disorder-induced SIT occurring at $\kappa_{c}\gg\kappa_{1}$~\citep{Feigelman_SIT_2010}.
Henceforth, the condition $\kappa\ll\kappa_{1}$ is assumed.

Hamiltonian~(\ref{eq:pseudospin-hamiltonian}) features minimal (gauge)
coupling to the discrete electromagnetic potentials $\phi_{i}$, $A_{i\rightarrow j}=-A_{j\rightarrow i}$.
As a consequence of the discrete nature of the model, these potentials
are defined on each site and on each \emph{directed} edge, respectively.
In the absence of external vector potential, the current operator
along a given edge $i\to j$ is given by 
\begin{equation}
I_{i\rightarrow j}=-c\,\frac{\partial H}{\partial A_{i\rightarrow j}}=8eD_{ij}\left(S^{x}_{i}S^{y}_{j}-S^{x}_{j}S^{y}_{i}\right).
\label{eq:current-op-def}
\end{equation}
Note that $I_{i\rightarrow j}$ is a four-particle operator in terms
of original electronic operators, expressing the fact that the charge
transport in the model occurs only because of the interaction. The
connection between $\phi_{i},\,A_{i\rightarrow j}$ and the real-space
electromagnetic potentials $\phi(\boldsymbol{r}),\,\boldsymbol{A}(\boldsymbol{r})$
is given by~\citep{superfluid-density-paper-2024} $\phi_{i}=\intop d^{3}\boldsymbol{r}|\psi_{i}(\boldsymbol{r})|^{2}\phi(\boldsymbol{r})$,
and $A_{i\rightarrow j}=\intop d^{3}\boldsymbol{r}\,\boldsymbol{\mathfrak{D}}_{i\rightarrow j}(\boldsymbol{r})\cdot\boldsymbol{A}(\boldsymbol{r})$,
where the field $\boldsymbol{\mathfrak{D}}_{i\rightarrow j}(\boldsymbol{r})$
is expressed in terms of variational derivatives of~$D_{ij}$ with
respect to external vector potential~\citep[Ch. 3]{Khvalyuk2025_thesis}.
The $\boldsymbol{\mathfrak{D}}$~field has the physical meaning of
the current density induced by tunneling of a Cooper pair from one
localized site to another, $\boldsymbol{j}(\boldsymbol{r})=I_{i\rightarrow j}\,\boldsymbol{\mathfrak{D}}_{i\rightarrow j}(\boldsymbol{r})$.
Importantly, charge conservation in real space implies~\citep{superfluid-density-paper-2024}
that $\left|\psi_{j}(\boldsymbol{r})\right|^{2}-\left|\psi_{i}(\boldsymbol{r})\right|^{2}=\nabla\cdot\boldsymbol{\mathfrak{D}}_{i\rightarrow j}(\boldsymbol{r})$.

The key quantity describing the low-frequency conductivity is the
retarded local current correlator $R_{ij}$:
\begin{equation}
R_{ij}(\omega)=(2e)^{2}\left\langle N_{ij}\right\rangle -\intop^{+\infty}_{0}dt\,ie^{i\omega t}\left\langle \left[I_{i\to j}(t),I_{i\to j}(0)\right]\right\rangle ,
\label{eq:current-correlator_def}
\end{equation}
where $N_{ij}=8eD_{ij}\left(S^{x}_{i}S^{x}_{j}+S^{y}_{j}S^{y}_{i}\right)$
is the appropriate diamagnetic term.

\paragraph{Solution by Belief Propagation. \label{sec:belief-propagation}}

To describe microscopic physical quantities, such as $R_{ij}$, we
employ the classical Belief Propagation~\citep{superfluid-density-paper-2024},
as suggested by the locally tree-like structure of the interaction
graph. This approach expresses local physical quantities in terms
of the eigenproblem of a certain two-spin Hamiltonian:
\begin{align}
H_{\left\langle ij\right\rangle } & =-\sum_{n=i,j}2\xi_{n}S^{z}_{n}-4D_{ij}\left(S^{x}_{i}S^{x}_{j}+S^{y}_{i}S^{y}_{j}\right)\nonumber \\
 & -\sum_{\alpha=x,y}\left(2h^{\alpha}_{i\rightarrow j}S^{\alpha}_{j}+2h^{\alpha}_{j\rightarrow i}S^{\alpha}_{i}\right).
\label{eq:two-sping_effective_Hamiltonian}
\end{align}
This Hamiltonian contains the local disorder $\xi_{i},\,\xi_{j}$
and the order-parameter fields $h_{i\rightarrow j},\,h_{j\rightarrow i}$,
which encode the local environment of the target edge~$\left\langle ij\right\rangle $.
The fields $h_{k\rightarrow i}$ are obtained by solving the self-consistency
equation for each \emph{directed} edge~$k\rightarrow i$:
\begin{equation}
h_{k\rightarrow i}=\sum_{j\in\partial i\backslash\left\{ k\right\} }D_{ij}\frac{h_{j\rightarrow i}}{B_{j\rightarrow i}}\,\tanh\frac{B_{j\rightarrow i}}{T},
\label{eq:h-equations}
\end{equation}
where $B_{j\rightarrow i}=\sqrt{\xi^{2}_{j}+h^{2}_{j\rightarrow i}}$,
and the sum over~$j$ runs over all neighbors of $i$ except~$k$.
Note that, due to the directedness of Eq.~(\ref{eq:h-equations}),
quantities with permuted vertex indices, e.g., $h_{i\to j}$ and $h_{j\to i}$,
are not equivalent.

Any physical quantity associated with a pair~$\left\langle ij\right\rangle $
of adjacent sites (see, e.g., Eq.~(\ref{eq:Im-R_particular-disorder-realization}))
is expressed through the eigenpairs $\left\{ E^{(n)}_{ij},\left|n_{ij}\right\rangle \right\} ,\,n=1,...,4$
of Hamiltonian~(\ref{eq:two-sping_effective_Hamiltonian}) with the
corresponding values of~$\left\{ \xi_{i},\,\xi_{j},\,h_{i\rightarrow j},\,h_{j\rightarrow i},\,D_{ij}\right\} $.
Using Eq.~(\ref{eq:h-equations}), one can express expectation values
of quantities on site~$j$ in a form that makes explicit the equivalence
and directedness of each edge incident to~$j$. An example is Eq.~(35)
of Ref.~\citep{superfluid-density-paper-2024} for the onsite order
parameter~$\Delta_{j}$, encoding the anomalous expectation~$\left\langle S^{-}_{j}\right\rangle $.
However, it is the set of fields $h_{i\to j}$ on each \emph{directed}
edge that encodes the complete statistical information, which is why
we focus exclusively on~$h_{i\to j}$. Moreover, it can be shown~\citep{superfluid-density-paper-2024}
that the statistics and physical properties of $h_{i\to j}$ closely
resemble those of~$\Delta_{j}$, justifying mild abuse of the term
``order parameter'' in reference to $h_{i\to j}$.

\paragraph{An approximate expression for $\text{Re}\sigma$ and theNM) Network
Model (NM).\label{sec:Re-sigma_via_average-dissipated-power}}

Applying a \emph{macroscopic} superconducting phase gradient $\nabla\varphi$
(e.g., as a boundary condition at the sample edges) creates a \emph{microscopic}
distribution of phase~$\varphi$ at each site, governed by the response
equations and charge conservation~\citep[Sec. IIIC]{superfluid-density-paper-2024}.
The real part of the conductivity is derived from the total Joule
heat: $P=\frac{1}{2}\intop d^{3}\boldsymbol{r}\,\text{Re}\sigma(\omega)\,\left|\boldsymbol{E}(\boldsymbol{r})\right|^{2}$,
where $\boldsymbol{E}(\boldsymbol{r})=-i\omega\left(\nabla\varphi/2e\right)$
is the external electric field. In terms of edge currents~$I_{i\to j}$,
the dissipated power is given by a sum of contributions from each
\emph{undirected} edge~$\left\langle jk\right\rangle $,
\begin{equation}
P=\frac{1}{2}\frac{1}{2e}\sum_{\left\langle jk\right\rangle }\text{Re}\left\{ -i\omega I^{*}_{k\to j}\left(\varphi_{j}-\varphi_{k}\right)\right\} .
\end{equation}
At frequency $\omega$ such that $\omega/\Delta_{0}\ll1$, the current
response can be represented as~\citep[Sec. IIIC]{superfluid-density-paper-2024}\footnote{At finite~$\omega>0$, Eq.~(\ref{eq:edge-current-response_functional-form})
also contains the nonlocal contributions from the phase differences
on edges~$e$ other than $\left\langle ij\right\rangle $; these
terms can be \emph{naively} estimated as~$(\omega/\Delta_{0})^{2d}$,
where $d$ is the distance between $e$ and $\left\langle ij\right\rangle $
on the graph, see~\citep[App. B]{superfluid-density-paper-2024}.} 
\begin{equation}
I_{i\to j}=\frac{1}{e}\,R_{ij}(\omega)\,\left(\varphi_{j}-\varphi_{i}\right),
\label{eq:edge-current-response_functional-form}
\end{equation}
where $R_{ij}$ is given by Eq.~(\ref{eq:current-correlator_def}). 

Moreover, at low frequencies, $R_{ij}$ is almost purely real, except
for \emph{rare instances} where the dissipative response contains
a resonance at sufficiently low frequencies, introducing a finite
imaginary contribution to the current from the first term of Eq. (\ref{eq:edge-current-response_functional-form}).
Since these instances are rare, the change in the distribution of
phases~$\varphi_{j}$ due to the finite imaginary part of $R_{ij}$
can be neglected, and one can use the phase distribution from the
$\omega=0$ case, where the response is purely superconducting. The
real part of the conductivity, $\text{Re}\sigma(\omega)$, is then
expressed as
\begin{equation}
\text{Re}\sigma(\omega)\approx n_{\text{e}}\,\frac{2}{\omega}\,\overline{\text{Im}R_{ij}(\omega)\,\frac{\left(\varphi_{j}-\varphi_{i}\right)^{2}}{\left(\overline{\nabla\varphi}\right)^{2}}}.
\label{eq:real-conductivity_via_local-response}
\end{equation}
Here, $n_{\text{e}}=n\left\langle K+1\right\rangle /2$ is the concentration
of undirected edges, the overline denotes averaging over all disorder
configurations, and $\varphi_{i}$ are the phases in the $\omega=0$
static problem with the mean phase gradient $\overline{\nabla\varphi}$
(see Ref.~\citep[Sec. IIIC]{superfluid-density-paper-2024}).

Eq.~(\ref{eq:real-conductivity_via_local-response}) provides an
approximate numerical method for computing the macroscopic real conductivity.
To this end, we employ an extended version of the protocol of Ref.~\citep[Sec. IIIC]{superfluid-density-paper-2024},
henceforth referred to as \emph{the network model}: \emph{i)}~generate
a large instance of a locally tree-like graph and disorder fields
$\xi_{i}$, \emph{ii)}~solve the self-consistency Eq.~(\ref{eq:h-equations})
for the order-parameter fields~$h_{i\to j}$ on each directed edge,
\emph{iii)}~compute the local responses $R_{ij}$ according to Eqs.~(\ref{eq:current-op-def})-(\ref{eq:two-sping_effective_Hamiltonian})
for each edge $\left\langle ij\right\rangle $, \emph{iv)}~numerically
solve the Kirchhoff equations for the superconducting phases $\varphi_{i}$
with a given phase difference $\varphi_{\text{right}}-\varphi_{\text{left}}=|\nabla\varphi|\times L$
in a geometry of a two-dimensional~\footnote{The use of the 2D geometry instead of the 3D one is a technical compromise
to achieve convergence of disorder averages. The qualitative results
of the analysis are expected to hold both in two and three dimensions
since the relevant dissipative processes are inherently microscopic.} brick of size~$L\times w$ (see Ref.~\citep[Sec. IIIC]{superfluid-density-paper-2024}
for details), and \emph{v)}~compute the required averages, such as
Eq.~(\ref{eq:real-conductivity_via_local-response}), using the resulting
large sample of $R_{ij}$ and $(\varphi_{j}-\varphi_{i})$. This procedure
is repeated for multiple disorder realizations to ensure a proper
disorder average.

\paragraph{Approximate analytical solution.\label{sec:approximate-analytical-solution}}

Directly computing the average in Eq.~(\ref{eq:real-conductivity_via_local-response})
requires the joint probability distribution of $\text{Im}R_{ij}(\omega)$
and $\left(\varphi_{j}-\varphi_{i}\right)^{2}$. This distribution
is only accessible via the numerical solution of the~NM. Remarkably,
the following approximate relation holds:
\begin{equation}
\text{Re}\sigma\left(\omega\ll\Delta_{0}\right)\approx\frac{2\eta\,n_{\text{e}}\overline{\left(\boldsymbol{r}_{i}-\boldsymbol{r}_{j}\right)^{2}}/\mathcal{D}}{\omega}\,\overline{\text{Im}R_{ij}(\omega)},
\label{eq:low-freq-conductivity_via_average-imaginary-response}
\end{equation}
where $\overline{\left(\boldsymbol{r}_{i}-\boldsymbol{r}_{j}\right)^{2}}=\left[\mathcal{D}/(\mathcal{D}+2)\right]r^{2}_{\text{loc}}$
for the present model in $\text{\ensuremath{\mathcal{D}}}$ spatial
dimensions, and the dimensionless coefficient~$\eta$ is nearly independent
of frequency. This relation is especially striking given that both
its sides are steep functions of frequency, as will be shown below.
The qualitative reason behind Eq.~(\ref{eq:low-freq-conductivity_via_average-imaginary-response})
is that $\text{Re}\sigma$ is dominated by the density of low-energy
excitations:
\begin{equation}
\text{Re}\sigma(\omega)\sim\overline{\delta\left(\omega-\Omega_{ij}\right)},\,\,\,\Omega_{ij}=\min_{n\neq m}\left|E^{(n)}_{ij}-E^{(m)}_{ij}\right|,
\label{eq:sigma-estimation_via-lowest-excitation-frequency}
\end{equation}
where $E^{(n)}_{ij}$ are the eigenenergies of $H_{\left\langle ij\right\rangle },$Eq.~(\ref{eq:pseudospin-hamiltonian}),
and $\Omega_{ij}$ is the minimal transition frequency for a given
edge~$\left\langle ij\right\rangle $. As will be shown below, the
spectral density of~$\Omega_{ij}$ exhibits a steep exponential dependence
on frequency. On the other hand, the average of the squared current
matrix element and phase difference in Eq.~(\ref{eq:real-conductivity_via_local-response})
carries at most a weak power-law dependence on~$\omega$; see \appref{End-Matter}
for details.

One further expects that the main temperature dependence is reproduced
in Eq.~(\ref{eq:low-freq-conductivity_via_average-imaginary-response}).
Indeed, finite temperature causes only a small change in the superfluid
stiffness, $\delta\Theta/\Theta\ll1$~\citep{superfluid-density-paper-2024},
suggesting a similarly small change in the phase differences $\varphi_{j}-\varphi_{i}$
and, consequently, in the value of $\eta$ in Eq.~(\ref{eq:low-freq-conductivity_via_average-imaginary-response}),
viz. $\delta\eta/\eta\sim\delta\Theta/\Theta\ll1$. However, one cannot
fully exclude that the temperature-dependent part of the correlations
between $\text{Im}R_{ij}$ and $(\varphi_{j}-\varphi_{i})$ is much
more pronounced among the strongly dissipating edges. A discussion
of this aspect is also presented in \appref{End-Matter}.

In addition, the numerical solution of the NM unambiguously demonstrates
a noticeable $\kappa$~dependence of $\eta$ (see \appref{End-Matter}).
Because of the approximate character of Eq.~(\ref{eq:low-freq-conductivity_via_average-imaginary-response}),
we do not conduct a detailed numerical analysis of this dependence.
However, the discussion of the $\omega,\,T$ dependencies of $\text{Re}\sigma$
remains qualitatively valid.

\begin{figure*}[t]
\begin{centering}
\includegraphics[width=0.9\columnwidth]{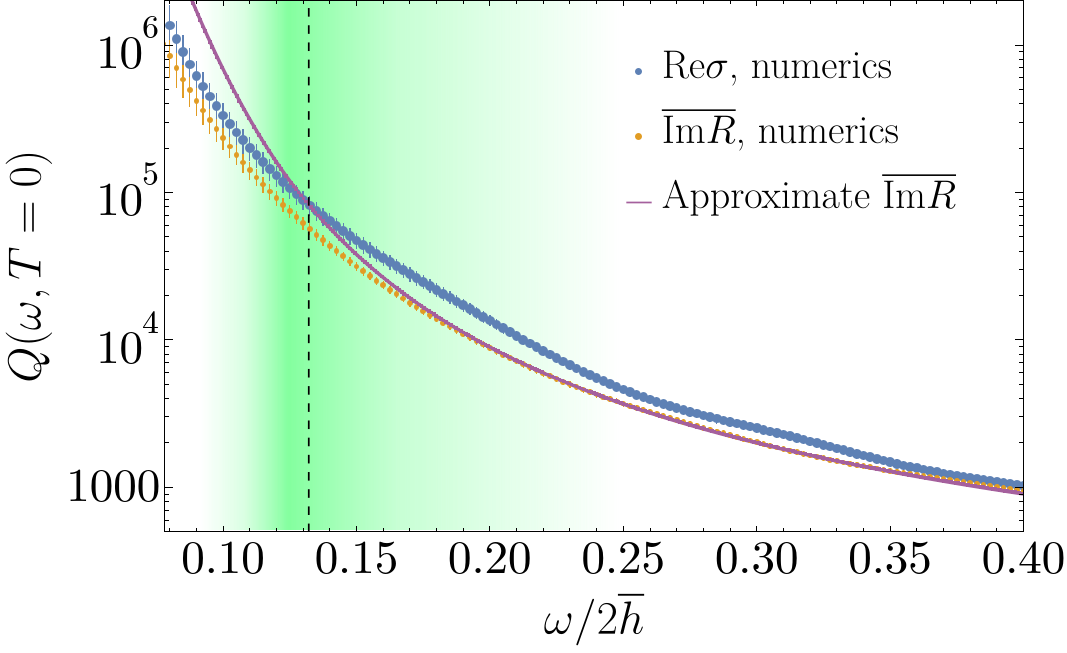}~~\includegraphics[width=0.864\columnwidth]{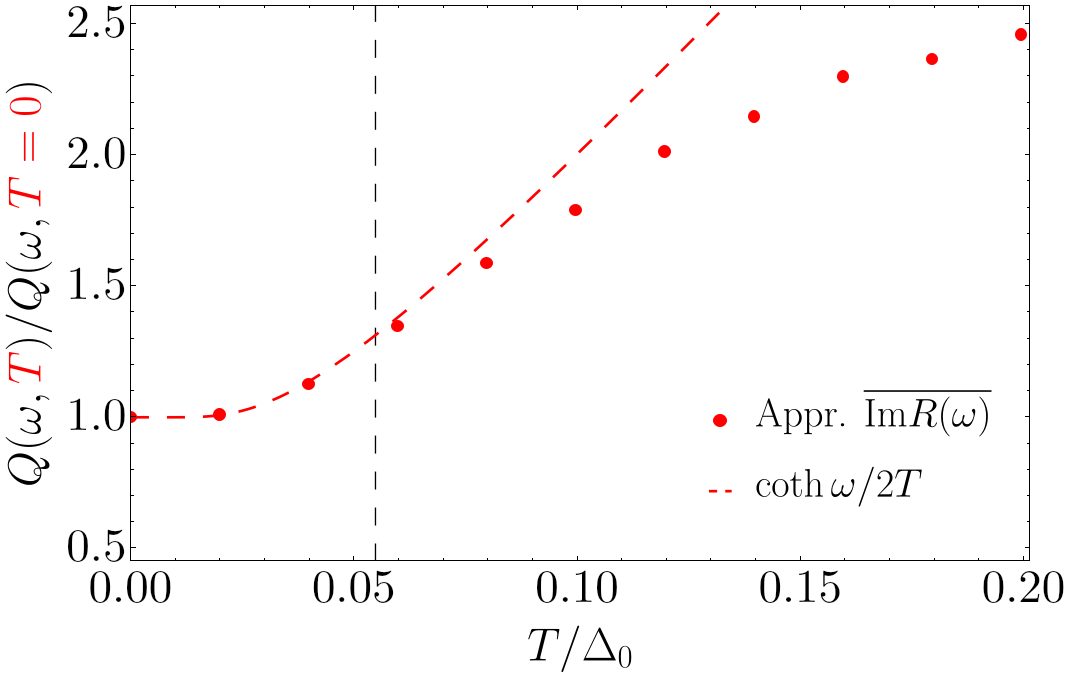}
\par\end{centering}
\caption{The behavior of the resonator quality factor $Q\propto\omega/\text{Re}\sigma$
in the pseudospin model with $K=10$, $\kappa=10$, $\lambda\approx0.1373$.
\emph{Left}:~dependence of $Q$ on frequency~$\omega$. Blue and
orange dots correspond to the data from $M=20$ realizations of the
NM of size $N\approx10^{6}$ with $r_{\text{loc}}/a=48.27$ (where
$a$ is the mean inter-site distance in real space), while the solid
purple line corresponds to Eq.~(\ref{eq:quality-factor_via_microscopic-response-statistics}),
with $P(h)$ found by population dynamics. To determine~$\eta$,
Eq.~(\ref{eq:low-freq-conductivity_via_average-imaginary-response})
was fitted on a broader frequency interval, $\omega/2\overline{h}\in\left[0.08,1\right]$,
causing low-frequency values of the approximation (orange) to be systematically
lower than the NM values (blue). The vertical dashed line corresponds
to $\omega=0.109\,\Delta_{0}$. The green region marks the range of
values that corresponds to the experimental data~\citep{charpentierFirstorderQuantumBreakdown2025,Charpentier2023_thesis},
with $T_{c}=1.4\,\text{K}$ and $\omega=3.85\,\text{GHz}$. The uncertainty
stems from the unaccounted-for quasiparticle suppression of the measured
$T_{c}$ relative to the model-predicted $T^{(0)}_{c}\approx2.4\,\overline{h}$~\citep{superfluid-density-paper-2024}.
The color intensity conveys the probability of a given~$\omega/2\overline{h}$.
\emph{Right:~}temperature dependence of $Q$, normalized to its value
at $T=0$, for $\omega=0.109\,\Delta_{0}$. Points correspond to Eq.~(\ref{eq:quality-factor_via_microscopic-response-statistics}),
with $P(h)$ found from population dynamics. The vertical dashed line
denotes $\omega=2T$. \label{fig:normalized-quality-factor_plot-theor}}
\end{figure*}

$\text{Im}R_{ij}(\omega>0)$ is expressed~\citep{superfluid-density-paper-2024}
in terms of the eigenpairs $\left\{ E^{(n)}_{ij},\left|n_{ij}\right\rangle \right\} $
of the two-spin Hamiltonian~(\ref{eq:two-sping_effective_Hamiltonian}):
\begin{equation}
\text{Im}R_{ij}(\omega)=\sum_{n\neq m}\left|I^{(mn)}_{ij}\right|^{2}W^{(nm)}_{ij}\,\frac{\pi}{2}\delta\left(\omega-\Omega^{(mn)}_{ij}\right),
\label{eq:Im-R_particular-disorder-realization}
\end{equation}
where $W^{(nm)}_{ij}=\left(e^{-E^{(n)}_{ij}/T}-e^{-E^{(m)}_{ij}/T}\right)/\sum_{m}e^{-E^{(m)}_{ij}/T}$,
$I^{(mn)}_{ij}=\left\langle m_{ij}\left|I_{i\rightarrow j}\right|n_{ij}\right\rangle $,
and $\Omega^{(mn)}_{ij}=E^{(m)}_{ij}-E^{(n)}_{ij}$. The average $\overline{\text{Im}R_{ij}(\omega)}$
is then obtained by averaging Eq.~(\ref{eq:Im-R_particular-disorder-realization})
over the ensemble of effective two-spin Hamiltonians obtained by sampling~$\left\{ \xi_{i},\,\xi_{j},\,h_{i\rightarrow j},\,h_{j\rightarrow i},\,D_{ij}\right\} $
(see Ref.~\citep[Sec. IIIE]{superfluid-density-paper-2024}). The
averaging procedure entails two technical complications: \emph{i)}~diagonalizing
of $H_{\left\langle ij\right\rangle }$ for each realization, and
\emph{ii)~}smoothing the $\delta$~function in $\text{Im}R_{ij}$.
The details of the associated numerical routine are presented in Ref.~\citep[App. G]{Khvalyuk2025_thesis}.

In the limit $\omega\ll\overline{h}$, where $\overline{h}$ is the
mean value of the order parameter, the averaging can be performed
analytically:
\begin{equation}
\frac{\overline{\text{Im}R_{ij}(\omega)}}{\tilde{R}}=\frac{\omega}{2\Delta_{0}}\tanh\frac{\omega}{2T}\intop^{\omega/2}_{0}dhP(h)\arccos\frac{2h}{\omega}.
\label{eq:local-response_via_order-parameter-distribution}
\end{equation}
Here, $\tilde{R}=(\pi\kappa/2)\left(2P_{0}\Delta_{0}\right)^{2}(2e)^{2}\overline{h}$,
and $P(h)$ stands for the probability density of the order parameter~$h_{i\to j}$.
This result expresses the fact that the relevant disorder configurations
are identical to those causing the temperature suppression of the
order parameter~\citep{superfluid-density-paper-2024}. The derivation
is detailed in Ref.~\citep[App. H]{Khvalyuk2025_thesis}; see also
\appref{End-Matter}.

Importantly, Eq.~(\ref{eq:local-response_via_order-parameter-distribution})
contains two sources of temperature dependence: \emph{i)}~the occupation
number $\tanh\{\omega/2T\}$ of the local mode, and \emph{ii)}~the
order-parameter distribution $P(h)$ that implicitly contains temperature.
The second aspect is important because Eq.~(\ref{eq:local-response_via_order-parameter-distribution})
is sensitive to the low-value tail of $P$, which, in turn, is strongly
temperature-dependent due to its steep profile~\citep{superfluid-density-paper-2024}.

From a technical point of view, Eqs.~(\ref{eq:low-freq-conductivity_via_average-imaginary-response})
and~(\ref{eq:local-response_via_order-parameter-distribution}) reduce
the computation of $\text{Re}\sigma$ (up to a frequency- and temperature-independent
constant) to the problem of finding~$P(h)$. This problem can be
efficiently addressed by \emph{population dynamics}, which amounts
to finding $P(h)$ such that the \emph{distributions} of left- and
right-hand sides of Eq.~(\ref{eq:h-equations}) are equal. Ref.~\citep[App. G]{Khvalyuk2025_thesis}
contains a brief review of the corresponding numerical routine, while
a more detailed exposition of the associated analytical techniques
can be found in Ref.~\citep{superfluid-density-paper-2024}.

\paragraph{Resonator quality factor as an experimental probe of the order-parameter
distribution.}

With the help of Eqs.~(\ref{eq:low-freq-conductivity_via_average-imaginary-response})
and~(\ref{eq:local-response_via_order-parameter-distribution}),
the inverse low-frequency quality factor of a resonator made of a
SDSC can be expressed as~\citep{charpentierFirstorderQuantumBreakdown2025,Charpentier2023_thesis}
\begin{equation}
\frac{1}{Q}=\frac{\text{Re}\ensuremath{\sigma}}{\text{Im}\sigma}=C\frac{\omega}{2\Delta_{0}}\tanh\frac{\omega}{2T}\intop^{\omega/2}_{0}dhP(h)\arccos\frac{2h}{\omega},
\label{eq:quality-factor_via_microscopic-response-statistics}
\end{equation}
where $\text{Im}\sigma=(2e)^{2}\Theta/(\omega d)$ is the imaginary
part of the conductivity, and $P(h)$ is the distribution of the order
parameter. The numerical coefficient $C=2\eta\,n\overline{\left(r_{i}-r_{j}\right)^{2}}d\,\left(2P_{0}\Delta_{0}\right)^{2}\frac{\Delta_{0}}{\Theta}\,(K+1)\left(\pi\kappa\overline{h}/12\Delta_{0}\right)$
is effectively independent of $T,\,\omega$ ($\Theta$ is the superfluid
stiffness, $d$ is the film's thickness). In Eq.~(\ref{eq:quality-factor_via_microscopic-response-statistics}),
one also neglects the weak dependence of the mean order parameter
$\overline{h}$ on temperature~\citep{superfluid-density-paper-2024}.
Eq.~(\ref{eq:quality-factor_via_microscopic-response-statistics})
constitutes the main result of this Letter.

\figref{normalized-quality-factor_plot-theor} shows the dependence
of $Q$ on both $\omega,\,T$ obtained by using Eq.~(\ref{eq:local-response_via_order-parameter-distribution})
for $\overline{\text{Im}R}$. The frequency dependence is compared
to the result of the numerical solution of the~NM. Two important
observations are in order: \emph{i)~$Q$ }decreases rapidly with
frequency due to the corresponding surge in the spectral density of
excitations. According to Eq.~(\ref{eq:local-response_via_order-parameter-distribution}),
this is a direct consequence of the steep profile~\citep{op-distribution-paper,superfluid-density-paper-2024}
of the order-parameter distribution~$P(h)$. \emph{ii)}~$Q$ initially
grows with temperature, reflecting thermal activation of the local
degrees of freedom, corresponding to the $\tanh\{\omega/2T\}$ factor
in Eq.~(\ref{eq:local-response_via_order-parameter-distribution}).
However, this trend is later slowed down by the increase in the spectral
density of excitations. Within Eq.~(\ref{eq:quality-factor_via_microscopic-response-statistics}),
this results from the growth~\citep{superfluid-density-paper-2024}
of the low-value tail of the order-parameter distribution.

Importantly, at the lowest frequencies, the \emph{left-hand} panel
of \figref{normalized-quality-factor_plot-theor} displays a discrepancy
between the NM and Eq.~(\ref{eq:quality-factor_via_microscopic-response-statistics}).
This difference originates from the fact that the NM obtains the order
parameter as the solution to the self-consistency Eq.~(\ref{eq:h-equations})
on a graph with a finite~$m_{\text{tree}}$, whereas Eq.~(\ref{eq:quality-factor_via_microscopic-response-statistics})
uses population dynamics to restore $P(h)$ and thus corresponds~\citep{superfluid-density-paper-2024}
to the limit~$m_{\text{tree}}\to\infty$. Therefore, the discrepancy
in \figref{normalized-quality-factor_plot-theor} signals the breakdown
of Belief Propagation for low $h$~values, as they acquire long-distance
correlations, despite our best efforts~\footnote{The model parameters for the data in \figref{normalized-quality-factor_plot-theor},
\emph{left} were carefully chosen to maximize $m_{\text{tree}}$ while
still being within the range of applicability of the employed numerical
methods, as detailed in Ref.~\citep[Ch. 5]{Khvalyuk2025_thesis}.}. Our findings are thus valid for frequencies above the problematic
range, whereas the qualitative picture within this range should be
studied more carefully.

Eq.~(\ref{eq:quality-factor_via_microscopic-response-statistics})
and the steep profile of $Q(\omega)$ can be exploited to experimentally
probe the tail of the order-parameter distribution. Namely, Eq.~(\ref{eq:local-response_via_order-parameter-distribution})
can be solved for~$P(h)$, rendering
\begin{equation}
P(h)\propto\frac{d}{dh}\intop^{2h}_{0}\frac{d\omega}{\sqrt{1-(\omega/2h)^{2}}}\,\frac{d}{d\omega}\frac{\coth\{\omega/2T\}}{\omega\,Q(\omega)}.
\label{eq:order-parameter-distribution_via_quality-factor}
\end{equation}
Eq.~(\ref{eq:order-parameter-distribution_via_quality-factor}) suggests
the following protocol for restoring $P\left(h\ll\Delta_{0}\right)$:
\emph{i)~}measure the low-frequency quality factor for a sequence
of plasmonic resonances on a thin strip resonator at low temperature,\emph{
ii)~}calculate the integrand by differentiating a numerical interpolation
of the data, \emph{iii)~}calculate the cumulative distribution function
$F(h)=\intop^{h}_{0}dhP(h)$ by means of numerical integration, and
\emph{iv)~}use one more round of numerical differentiation to restore~$P(h)$
up to an overall factor. The result can further be compared to previous
indirect experimental~\citep{Kamlapure12,Sacepe_2011_for-pair-preformation}
and numerical~\citep{Bouadim11,Lemarie2013,Trivedi2021} probes of
the same quantity.

%% file: _Text__Discussion.tex
\paragraph{Conclusions.\label{sec:Conclusions}}

We analyzed the real part of conductivity, $\text{Re}\sigma(\omega)$,
in SDSCs. We demonstrated that $\text{Re}\sigma$ at low frequency
is dominated by bulk collective localized excitations. These excitations
reside in the regions of local suppression of the superconducting
order parameter and are thus directly linked to the intrinsic inhomogeneity
of the superconducting state. The sharp profile of the spectral density
of these excitations translates into a steep increase of $\text{Re}\sigma$
with frequency, roughly following the low-value tail of the order-parameter
distribution,~$P(h\ll\Delta_{0})$ {[}see Eq.~(\ref{eq:quality-factor_via_microscopic-response-statistics}){]}.

Our results are in qualitative agreement with the recent experimental
data~\citep{charpentierFirstorderQuantumBreakdown2025,Charpentier2023_thesis}
on resonators made of $\mathrm{InO}_{x}$. This includes both the
overall magnitude of the internal resonator quality factor $Q\propto1/\text{Re}\sigma$
and its temperature dependence (see \figref{normalized-quality-factor_plot-theor}).
 To test relation~(\ref{eq:order-parameter-distribution_via_quality-factor})
between $Q(\omega)$ and $P(h)$, a detailed low-temperature measurement
of $Q(\omega)$ for an SDSC-based resonator is desirable, e.g., using
the technique of Refs.~\citep{charpentierFirstorderQuantumBreakdown2025,Charpentier2023_thesis}.

\figref{normalized-quality-factor_plot-theor}, \emph{left,} suggests
improving coherence times of SDSC-based quantum devices by lowering
the operating frequency. Because $P(h)$ is steep, halving $\omega$
can raise $Q$ by an order of magnitude for the same film, provided
the film is sufficiently disordered for the present theory to apply.
This improvement is practically important for several qubit designs~\citep{Mooij05,Mooij06,kermanMetastableSuperconductingQubit2010,Astafiev12,DeGraaf18,peltonenHybridRfSQUID2018,Bylander19},
advancing scalable quantum computing.

Our results raise a number of physical questions beyond the technical
issues of the employed simplifications. First, our analysis does
not yield the spatial structure of the localized collective modes
in question. Second, these excitations have been conjectured~\citep{superfluid-density-paper-2024}
to be responsible for the near-power-law suppression of the superfluid
stiffness~$\Theta$ as a function of~$T$. However, directly computing
$\Theta(T)-\Theta(0)$ via the Ferrell-Glover-Tinkham relation~\citep{ferrell1958_supercond-sum-rule}
is hindered because the corresponding integral converges at frequencies
comparable to the superconducting energy scales, whereas our theory
is limited to much lower frequencies.  Finally, the additional low-temperature
entropy due to the discussed collective modes appears important~\citep{charpentierFirstorderQuantumBreakdown2025}
for the structure of the phase diagram of SDSCs near the disorder-driven
quantum phase transition. At the same time, Ref.~\citep{poboikoMeanfieldTheoryFirstorder2024}
demonstrates that Coulomb repulsion is essential for describing the
first-order nature of this transition. A consistent treatment incorporating
both ingredients will be presented elsewhere.

Some of the aforementioned experimental features were also observed~\citep{GrAl,Rotzinger-grAl2023}
in high-resistance granular aluminum films. However, the electron
(near) localization in granular materials arises from their fine-grained
structure, rather than from the single-particle Anderson localization~\citep{Ma_Lee_1985_Ref-to-pseudospins,Feigelman_Fractal-SC_2010}
relevant to the present work. Nevertheless, the presence of localization
effects in both types of systems suggests that similar low-frequency
dissipation mechanisms may be operative. 

\begin{acknowledgments}
The authors would like to thank Denis Basko, Thibault Charpentier
and Benjamin Sacépé for numerous fruitful discussions. A.V.K. is grateful
for the support by Laboratoire d’excellence LANEF in Grenoble (No.
ANR-10-LABX-51-01).
\end{acknowledgments}

%% file: _App__End-Matter.tex
\section{End Matter\label{app:End-Matter}}

\paragraph{Quality of the approximation for $\text{Re}\sigma(\omega)$.}

Relation~(\ref{eq:real-conductivity_via_local-response}) represents
an empirical shortcut to the solution of the NM as it replaces all
possible statistical correlations between $\text{Im}R_{ij}(\omega')$
and $\left(\varphi_{j}-\varphi_{i}\right)^{2}$ with a single coefficient~$\eta$.
To characterize the quality of this approximation, one considers the
following generalization of this coefficient:
\begin{equation}
\eta(\omega):=\frac{\overline{\intop^{\omega}_{0}d\omega'\,\text{Im}R_{ij}(\omega')\,\left(\varphi_{j}-\varphi_{i}\right)^{2}}}{\,\overline{\intop^{\omega}_{0}d\omega'\,\text{Im}R_{ij}(\omega')}\times\left(\overline{\nabla\varphi}\right)^{2}\overline{\left(\boldsymbol{r}_{i}-\boldsymbol{r}_{j}\right)^{2}}/\mathcal{D}},
\label{eq:eta-integral-coeff}
\end{equation}
where $\overline{\left(\boldsymbol{r}_{i}-\boldsymbol{r}_{j}\right)^{2}}=\left[\mathcal{D}/\left(\mathcal{D}+2\right)\right]r^{2}_{\text{loc}}$
for the simple pseudospin model in $\mathcal{\mathcal{D}}$ spatial
dimensions, $\overline{\nabla\varphi}$ is the mean external phase
gradient, and all other averages $\overline{\bullet}$ are estimated
numerically from the solution of the $\omega=0$ NM. The integration
in Eq.~(\ref{eq:eta-integral-coeff}) is needed to facilitate the
averaging procedure, since expression~(\ref{eq:Im-R_particular-disorder-realization})
for $\text{Im}R_{ij}$ contains a $\delta$~function of frequency,
an inconvenient object for numerical estimations.

The numerator of Eq.~(\ref{eq:eta-integral-coeff}) is proportional
to the integral $\intop^{\omega}_{0}d\omega'\,\omega'\text{Re}\sigma(\omega')$
as found by the solution to the NM itself, Eq.~(\ref{eq:real-conductivity_via_local-response}),
while the denominator of Eq.~(\ref{eq:eta-integral-coeff}) represents
the same integral of the approximate expression~(\ref{eq:low-freq-conductivity_via_average-imaginary-response}).
Should Eq.~(\ref{eq:low-freq-conductivity_via_average-imaginary-response})
be a faithful representation of the dissipative conductance, $\eta(\omega)$
will be a constant function of frequency, while the actual $\omega$~dependence
of $\eta$ characterizes the inaccuracy of the approximation.

\begin{figure}
\begin{centering}
\includegraphics[width=1\columnwidth]{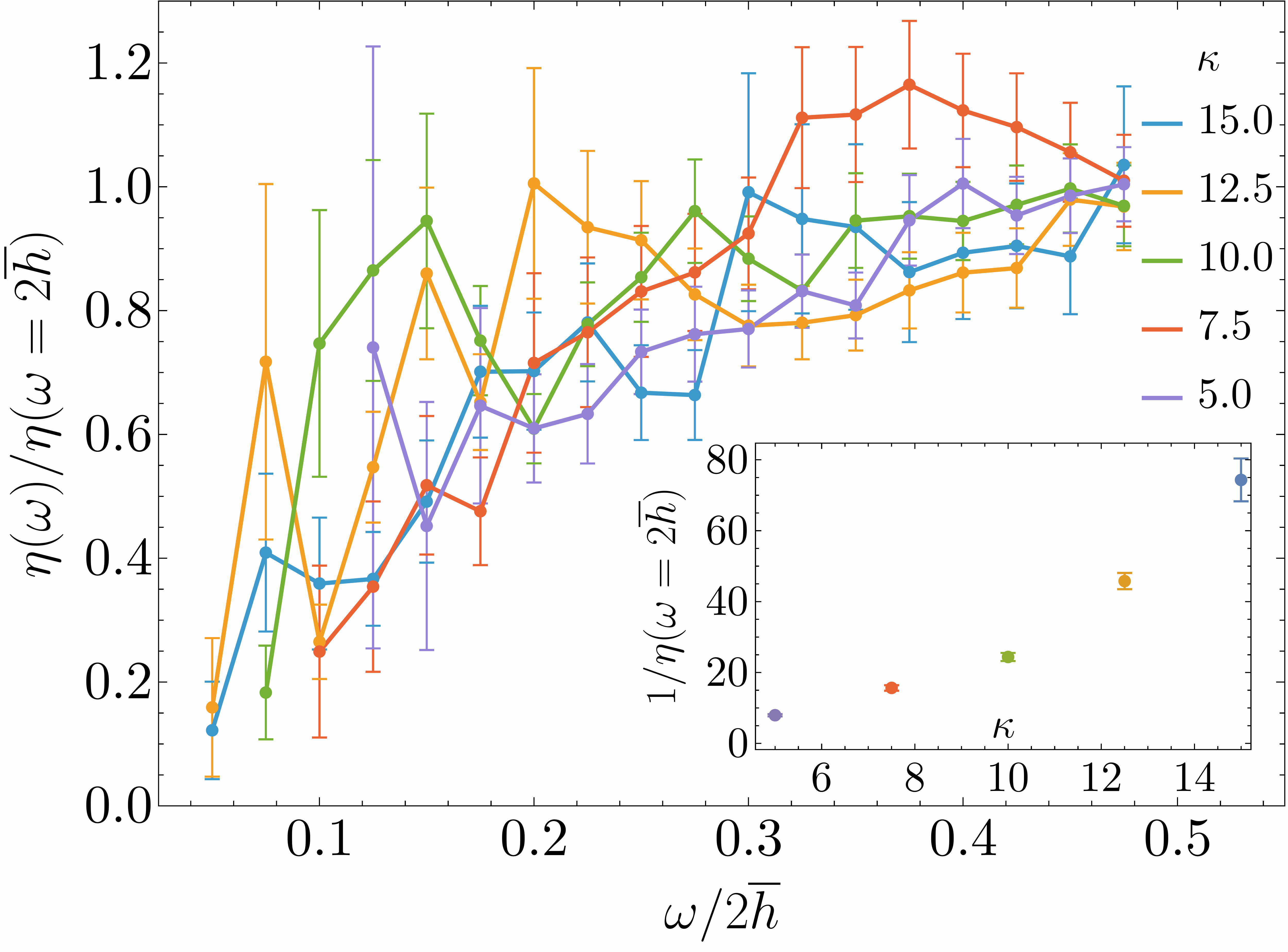}
\par\end{centering}
\caption{$\eta(\omega)$, defined in Eq.~(\ref{eq:eta-integral-coeff}),
for $\mathcal{D}=2$-dimensional pseudospin model as a function of
$\omega$ for various disorders. The parameters and system sizes are
the same as in \figref{normalized-quality-factor_plot-theor}, apart
from~$\lambda$ used to tune~$\kappa$. For each curve in the main
panel, $\omega$ and $\eta(\omega)$ are normalized to, respectively,
$2\overline{h}$ and $\eta(\omega=2\overline{h})$ for the given $\kappa$~value.
The inset shows the evolution of $1/\eta(\omega=2\overline{h})$ with
disorder. The error bars correspond to statistical fluctuations of
both the numerator and denominator in Eq.~(\ref{eq:eta-integral-coeff})
due to both the finite number of disorder realizations and the finite
size of each disorder realization. The number of disorder realizations
for each~$\kappa$ varied from $2$ to $20$ to achieve comparable
error bars. In the main panel, only the points with the relative error
below 100\% are presented, while the low-$\omega$ data of poor quality
are left out.\label{fig:integrated-frequency-dependence_various-kappa}}
\end{figure}

The resulting curves for $\eta(\omega)$ in $\mathcal{D}=2$ dimensions
are shown in \figref{integrated-frequency-dependence_various-kappa}
for various levels of disorder. As the main panel illustrates, $\eta(\omega)$
exhibits a notable frequency dependence, thus demonstrating the approximate
character of Eq.~(\ref{eq:low-freq-conductivity_via_average-imaginary-response}).
Moreover, the inset in \figref{integrated-frequency-dependence_various-kappa}
demonstrates that the typical value of $\eta$ depends significantly
on the dimensionless disorder strength $\kappa$. For these reasons,
the value of $\eta$ for \figref{normalized-quality-factor_plot-theor},~\emph{left}
was found by averaging the relation of the two sides of Eq.~(\ref{eq:low-freq-conductivity_via_average-imaginary-response})
over a range of frequencies $\omega\in\left[0.16\overline{h},2\overline{h}\right]$.
The technical procedure for computing the corresponding averages is
described in Ref.~\citep[App. G]{Khvalyuk2025_thesis}, and \citep[Ch. 5]{Khvalyuk2025_thesis}
contains a more detailed discussion of the $\eta(\omega)$ dependence
and its origins.

\paragraph{Approximate expression for $R_{ij}(\omega)$ at dissipative edges.}

The condition of low excitation frequency, $\Omega_{ij}\ll\overline{h}$,
is rather restrictive for possible disorder configurations of the
effective two-spin Hamiltonian, Eq.~(\ref{eq:two-sping_effective_Hamiltonian}).
In Ref.~\citep[App. H]{Khvalyuk2025_thesis} it is shown that two
conditions have to be met: \emph{i)~}one of the two $\xi$ fields---without
loss of generality, let this be the field $\xi_{1}$ on the first
spin---has to be the largest scale, $\left|\xi_{1}\right|\gg\left|\xi_{2}\right|,\,D_{12},\,h_{1\to2},\,h_{2\to1},\,\omega$,
and \emph{ii)}~both local fields of the other spin have to be of
the order of frequency: $\left|\xi_{2}\right|,h_{1\to2}\apprle\omega$.
Only under these conditions does the Hamiltoniani~(\ref{eq:two-sping_effective_Hamiltonian})
posses an excitation with frequency $\Omega_{12}=\omega\ll\overline{h}$.
Under the same conditions, direct perturbation theory in powers of
$1/\left|\xi_{1}\right|$ yields the following result for the local
superfluid response, Eq.~(\ref{eq:current-correlator_def}):
\begin{align}
 & \omega=\Omega_{12}\approx2\sqrt{\xi^{2}_{2}+\left(h_{2\to1}+D_{12}h_{1}/\left|\xi_{1}\right|\right)^{2}},
\label{eq:excitation-frequency_approx-expression}\\
 & \frac{\text{Im}R_{12}(\omega)}{(2e)^{2}}\approx\left(D_{12}\frac{h_{1}}{\left|\xi_{1}\right|^{2}}\right)^{2}\tanh\frac{\Omega_{12}}{2T}\,\pi\delta\left(\omega-\Omega_{12}\right),
\label{eq:Im-R_low-freq_approx-expression}\\
 & \frac{R_{12}}{(2e)^{2}}\approx\frac{2D_{12}\,h_{1\to2}\,h_{2\to1}}{\Omega_{12}\,\left|\xi_{1}\right|}\tanh\frac{\Omega_{12}}{2T}.
\label{eq:superlfuid-response_approx-expression}
\end{align}
Eq.~(\ref{eq:excitation-frequency_approx-expression}) further implies
the corresponding smallness of one of the two order parameter fields,
$h_{1\to2}\apprle\omega/2\ll\overline{h}$, explaining the connection
of these excitations to the low-value tail of the order parameter.
Subsequent averaging of Eq.~(\ref{eq:Im-R_low-freq_approx-expression})
over disorder renders Eq.~(\ref{eq:local-response_via_order-parameter-distribution}).

\paragraph{Statistical correlation between $\left(\varphi_{i}-\varphi_{j}\right)^{2}$
and $R_{ij}(\omega)$.}

\begin{figure*}
\begin{centering}
\includegraphics[width=1\textwidth]{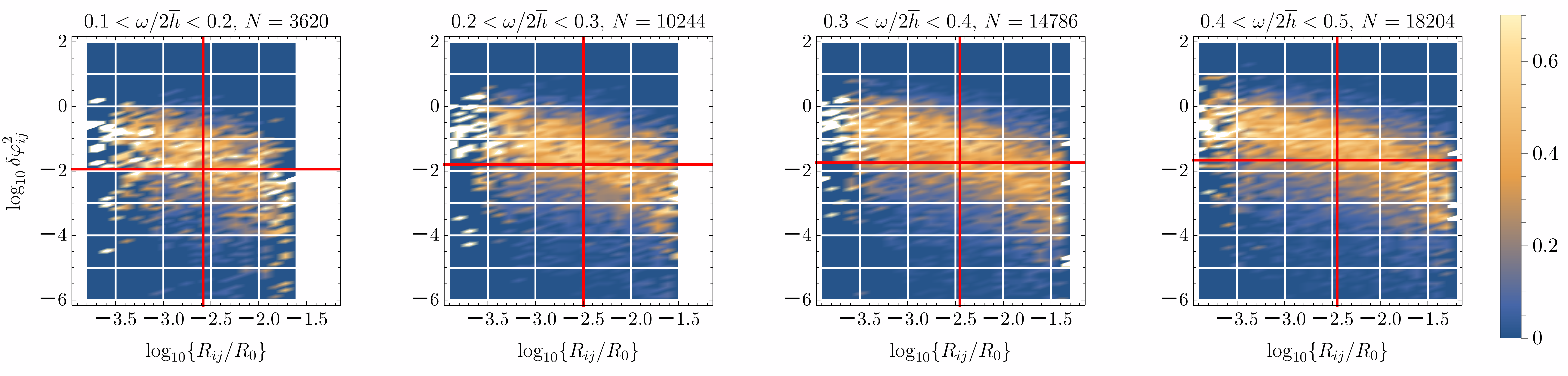}
\par\end{centering}
\caption{A series of color plots of the conditional probability $P\left(\log_{10}\delta\varphi^{2}_{ij}\,|\,\log_{10}R_{ij}/R_{0};\,\omega_{1},\omega_{2}\right)$,
defined in Eq.~(\ref{eq:phase-conditional-probability-expression}),
for various frequency intervals $\left[\omega_{1},\omega_{2}\right]$,
specified on top of each plot in units of $2\overline{h}$. $R_{ij}$
is measured in units of $R_{0}=\left(2P_{0}\Delta_{0}\right)^{2}(2e)^{2}\Delta_{0}$.
The red lines indicate the averages $\overline{\log_{10}\delta\varphi^{2}_{ij}}$,
$\overline{\log_{10}R_{ij}/R_{0}}$ conditioned on the respective
frequency interval, $\omega_{1}<\Omega_{ij}<\omega_{2}$. The histogram
is constructed from the dataset used in \figref{normalized-quality-factor_plot-theor},
with the histogram bin sizes $\delta\log_{10}R/R_{0}=0.1$, $\delta\log_{10}\delta\varphi^{2}_{ij}=0.1$.
The number of edges $N$ contributing to each histogram is specified
at the top of the respective plot. The irregularity of the plot on
both sides of the $\log_{10}R_{ij}/R_{0}$ range is a finite size
effect due to small marginal probability $P(\log_{10}R_{ij}/R_{0}|\omega_{1},\omega_{2})$.
 \label{fig:l-squared_conditioned-on-R-omega}}
\end{figure*}

In this section, we review the correlations between the local superfluid
response of a given edge $\text{Re}R_{ij}(\omega=0)$, studied in
Ref.~\citep{superfluid-density-paper-2024} and denoted here $R_{ij}$
for brevity, and the superconducting phase difference on the same
edge. We focus our analysis on the edges that have sufficiently low
excitation frequency to contribute to $\text{Re}\sigma$ and thus
influence its temperature dependence. To this end, \figref{l-squared_conditioned-on-R-omega}
visualizes the following conditional probability density
\begin{align}
 & P\left(\psi\,|\,\rho;\,\omega_{1},\omega_{2}\right):=\nonumber \\
 & \left.\overline{\delta\left(\log_{10}\delta\varphi^{2}_{ij}-\psi\right)\delta\left(\log_{10}R_{ij}/R_{0}-\rho\right)}_{\omega_{1}<\Omega_{ij}<\omega_{2}}\right/\nonumber \\
 & \overline{\delta\left(\log_{10}R_{ij}/R_{0}-\rho\right)}_{\omega_{1}<\Omega_{ij}<\omega_{2}}.
\label{eq:phase-conditional-probability-expression}
\end{align}
Here, $R_{0}=\left(2P_{0}\Delta_{0}\right)^{2}(2e)^{2}\Delta_{0}$,
$\Omega_{ij}$ is the lowest excitation frequency of edge $ij$ {[}see
Eq.~(\ref{eq:sigma-estimation_via-lowest-excitation-frequency}){]},
the subscript means that only the edges satisfying $\omega_{1}<\Omega_{ij}<\omega_{2}$---and
thus contributing to $\text{Re}\sigma$ in the same frequency range---are
used in the averaging, and $\delta\varphi^{2}_{ij}$ is the normalized
squared phase difference, $\delta\varphi^{2}_{ij}=\mathcal{D}\left[\left(\varphi_{i}-\varphi_{j}\right)/\overline{\nabla\varphi}\right]^{2}\left/\overline{\left(\boldsymbol{r}_{i}-\boldsymbol{r}_{j}\right)^{2}}\right..$

\figref{l-squared_conditioned-on-R-omega} reveals the main qualitative
features of the joint statistics of $R_{ij}$ and $\delta\varphi^{2}_{ij}$
for the dissipative edges: \emph{i)~}The marginal distribution of
\emph{the logarithm of} $\delta\varphi^{2}_{ij}$, $P\left(\psi|\omega_{1},\omega_{2}\right)=\intop^{\infty}_{-\infty}d\rho\,P\left(\psi|\rho,\omega_{1},\omega_{2}\right)$,
is broad, with the values of $\delta\varphi^{2}_{ij}$ distributed
across multiple decades. This creates difficulties for the numerical
analysis and, in particular, explains the strong statistical fluctuations
observed in \figref{integrated-frequency-dependence_various-kappa}.
\emph{ii)}~$R_{ij}$ and $\delta\varphi^{2}_{ij}$ are substantially
anticorrelated, with larger $R_{ij}$ leading to smaller $\delta\varphi^{2}_{ij}$.
\emph{iii)}~These correlations are only weakly sensitive to the frequency
interval.

These features are essential for understanding the correct temperature
dependence of the dissipative conductivity~$\text{Re}\sigma$. Indeed,
as one increases the temperature starting from $T=0$, the change
in $\text{Re}\sigma$ originates from both $\text{Im}R_{ij}$ and
$\left(\varphi_{i}-\varphi_{j}\right)^{2}$, according to Eq.~(\ref{eq:real-conductivity_via_local-response}).
The first of these two factors is correctly captured by the approximate
Eq.~(\ref{eq:low-freq-conductivity_via_average-imaginary-response}).
On the other hand, the temperature effect of $\left(\varphi_{i}-\varphi_{j}\right)^{2}$
is hard to analyze numerically, as it requires solving the $\omega=0$
NM for every~$T$, whereas all NM data in this work correspond to
$T=0$ and have already required substantial computational time. However,
\figref{l-squared_conditioned-on-R-omega} suggests that the temperature
shift of $\left(\varphi_{i}-\varphi_{j}\right)^{2}$ \emph{for the
low-frequency edges} can be inferred from that of~$R_{ij}$. According
to Eq.~(\ref{eq:superlfuid-response_approx-expression}), the main
temperature dependence of both $R_{ij}$ and $\overline{\text{Im}R_{ij}(\omega)}$
is set by $\Omega_{ij}$ via the common $\tanh\frac{\omega}{2T}$
factor, whereas the typical temperature scale for the change of $\Omega_{ij}$
and the $h$~fields is~$\overline{h}$, which is assumed to be much
higher than~$\omega$. Therefore, one expects $R_{ij}$ to diminish
strongly with $T$ for the same low-frequency edges that contribute
to dissipation. This conclusion is then transferred to $\delta\varphi^{2}_{ij}$
by its anticorrelation with $R_{ij}$, implying additional temperature
dependence of $\text{Re}\sigma$ that is not captured by Eq.~(\ref{eq:low-freq-conductivity_via_average-imaginary-response}).
Ref.~\citep[Ch. 5]{Khvalyuk2025_thesis} conducts further empirical
analysis of these correlations, which are argued to only alter the
quantitative shape of the temperature dependence, while preserving
the qualitative behavior shown in \figref{normalized-quality-factor_plot-theor}.
A more detailed study of the temperature dependence of $\text{Re}\sigma$
is a subject of future work.